\documentclass[acmlarge, nonacm]{acmart}

\AtBeginDocument{%
  \providecommand\BibTeX{{%
    \normalfont B\kern-0.5em{\scshape i\kern-0.25em b}\kern-0.8em\TeX}}}

\setcopyright{acmcopyright}
\copyrightyear{2020}
\acmYear{2020}
\acmDOI{10.1145/1122445.1122456}

\usepackage{tikz}
\usepackage{booktabs}
\usepackage[resetlabels]{multibib}
\newcites{W}{Appendix: Wearables Analyzed}
\usepackage{colortbl}
\usepackage{makecell}
\usepackage[normalem]{ulem}

\definecolor{safety}{RGB}{171,235,198}
\definecolor{connect}{RGB}{174,214,241}
\definecolor{health}{RGB}{250,215,160}
\definecolor{nurturing}{RGB}{235,222,240}

\definecolor{highlight}{RGB}{0,0,0}
\definecolor{EKC}{RGB}{217,95,2}
\definecolor{rachael}{RGB}{204,0,204}

\newcommand{\citew}[1]{\textsuperscript{\citeW{#1}}}

\usepackage{xparse}
\let\oriCitew\citew

\RenewDocumentCommand{\citew}{m}{%
  \renewcommand{\citenumfont}[1]{W##1}%
  \oriCitew{#1}%
  \renewcommand{\citenumfont}[1]{##1}%
}

\newcommand{\highlight}[1]{\textcolor{highlight}{#1}}

\begin{document}

\title{Investigating Opportunities to Support Kids' Agency and Well-being: A Review of Kids' Wearables}









\author{Rachael Zehrung}
\affiliation{\institution{University of Maryland, College Park}}
\email{rzehrung@umd.edu}

\author{Lily Huang}
\affiliation{\institution{University of Maryland, College Park}}
\email{lhuang12@umd.edu}

\author{Bongshin Lee}
\affiliation{\institution{Microsoft Research}}
\email{bongshin@microsoft.com}

\author{Eun Kyoung Choe}
\affiliation{\institution{University of Maryland, College Park}}
\email{choe@umd.edu}


\begin{abstract}
Wearable devices hold great potential for promoting children’s health and well-being. However, research on kids' wearables is sparse and often focuses on their use in the context of parental surveillance. To gain insight into the current landscape of kids' wearables, we surveyed 47 wearable devices marketed for children. We collected rich data on the functionality of these devices and assessed how different features satisfy parents’ information needs, and identified opportunities for wearables to support children’s needs and interests. We found that many kids' wearables are technologically sophisticated devices that focus on parents' ability to communicate with their children and keep them safe, as well as encourage physical activity and nurture good habits. We discuss how our findings could inform the design of wearables that serve as more than monitoring devices, and instead support children and parents as equal stakeholders, providing implications for kids’ agency, long-term development, and overall well-being. Finally, we identify future research efforts related to designing for kids’ self-tracking and collaborative tracking with parents. 

\end{abstract}



\maketitle

\section{Introduction}
\label{sec:introduction}

Kids' wearables have huge potential to impact kids' health \& well-being, self-regulation skills, and safety. Given that only one in four children of age 6 to 17 meet the recommended physical activity level~\cite{usreportcard2018}, fitness trackers have a potential to improve their physical activity level while introducing the importance of health tracking early on. Moreover, kids' wearables can be equipped with features to nurture good habits and daily routines (e.g., brushing teeth, keeping a consistent wake-up and bed time). They can also help ensure kids’ safety, for example, by allowing parents to track kids' whereabouts and to communicate with their kids. For parents who are hesitant to purchase a smartphone for early aged kids, wearables can be seen as an alternative smart device while meeting some of parents' and kids' needs. 

Recognizing these potentials, major fitness companies have launched kids' wearables (e.g., Garmin vivofit jr.\citew{garminvivofitjr2}\footnote{Throughout this article we refer to the wearables via a super-scripted, bracketed number to an item in APPENDIX.} in 2016, followed by Fitbit Ace\citew{fitbitace2} in 2018\footnote{Fitbit discontinued the Ace and replaced it with the Ace 2 in 2019.}) while small businesses introduced Kickstarter campaigns (e.g., Octopus\citew{octopuskickstarter} in 2017). \highlight{The specifications of kids' wearables vary drastically and are still evolving. In terms of the form factor, many kids' wearables are smartwatches or wristbands, but other devices are designed to be attached to backpacks and clothing (e.g., Jiobit\citew{jiobit}), or even embedded in clothing (e.g., Smart B'zT\citew{smartbzt}).} Although this is in part because the market is still at its early stage, it may indicate the diverse goals and needs of two different parties---the buyer (parent) and wearer (kid) of the device. 
\highlight{Previous work has introduced novel wearable prototypes for children (e.g.,~\cite{teh2008huggy, ryokai2014energybugs, kazemitabaar2016rewear}), but they are sparse and limited in their scope and form factor. Furthermore, research on kids' wearables has centered on Child Surveillance Technology (CST) and Child GPS Technology (CGT) \cite{bettany2016socio}, which are designed for parents who want to ensure the safety of their child. In contrast, commercial kids' wearables are widely available and have recently become a topic of interest in HCI and family informatics research \cite{oygur2020raising}. Existing commercial kids' wearables may also reflect technological trends and consumers' needs, and we aim to complement the previous research on kids' wearables by looking into how these commercial devices are intended to be used in the wild.} 
\highlight{More specifically, we set out to characterize the technological landscape of kids’ wearables and evaluate the extent to which existing commercial devices meet parents’ information needs, with the broader goal of investigating if and how kids’ wearables can be used to enhance children’s agency and well-being.}


To that end, we surveyed 47 wearable devices marketed for kids. In analyzing the features of these devices, we built upon Kuzminykh and Lank's work on parents' information needs~\cite{kuzminykh2019much}, which were based on interviews and ecological momentary assessment (EMA) studies with parents. As a starting point, we used the four categories drawn from their work---routine, health, schooling, and social \& emotional information needs---to analyze specific features embedded in existing kids wearables. We then revised and extended these categories because our analysis revealed features that can satisfy additional needs that are not covered in Kuzminykh and Lank. 
We found that kids' wearables were, in fact, wearables for \textit{caregivers}, designed mainly to ensure caregivers' peace of mind; the devices were equipped with sophisticated features to meet caregivers' safety and communication needs. That said, 57\% of devices offered unique features to support children's health and 38\% provided features to develop good habits, a promising direction that wearables for kids can offer beyond the current dominant trend toward surveillance. Such a shift in viewing kids as an object of monitoring to an agent of self-tracking affords designs for helping kids learn and practice \textit{self-tracking} early on and develop a sense of agency to manage their health and well-being. \highlight{We highlight opportunities to support children's self-tracking by designing for children's interests and values, to facilitate parental involvement through collaborative tracking and scaffolding, and to enrich parent-child communication through cross-device collaboration.}

\highlight{In summary, the key contributions of our work are: 
\begin{enumerate}
    \item surveying and synthesizing the technological capabilities of existing commercial kids’ wearables; 
    \item evaluating the extent to which existing capabilities of kids’ wearables meet parents’ information needs; and 
    \item identifying design and research opportunities for kids’ wearables that strike the balance between children’s self-regulation and parents’ information needs. 
\end{enumerate}
}


\section{Related Work}
\label{sec:related-work}
In this section, we cover related work on self-tracking with wearable devices, child surveillance technologies, and tracking within a family context.

\subsection{Wearable Devices for Health}
Wearables allow people to track different types of data about themselves, such as physical activity (e.g.,~\cite{morris2014recofit, fritz2014persuasive, zhao2017keeping, henriksen2018using}), sleep (e.g., \cite{daskalova2018investigating, pollak2001accurately, ko2015consumer}), eating (see ~\cite{bell2020automatic} for an overview), and stress (see~\cite{gradl2019overview} for an overview). Commercial devices from major companies, such as Apple, Samsung, Fitbit, and Garmin have become more popular to the point that 21\% of American adults regularly used a smartwatch or fitness tracker in 2019~\cite{pew2020}. When used as fitness interventions, wearables have been shown to promote physical activity and weight loss~\cite{coughlin2016use}, though that is not always the case \cite{jakicic2016effect}. 


Recognizing wearables' potential as communication devices, several telecommunications companies created devices for children (e.g., Verizon GizmoWatch 2 \citew{verizongizmowatch2}) that enable calling and texting alongside GPS tracking. That said, activity tracking has become a growing focus for children's devices as well. For example, Garmin \citew{garminvivofitjr2} and Fitbit \citew{fitbitace2} introduced their own versions of activity trackers for children. While these devices for kids are designed to promote general physical activity and healthy habits, they are not equipped with features to infer calories burned and heart rate, which are commonly observed in adult activity trackers. In addition, children's wearables are generally more colorful and have more features to encourage usage of the device itself (e.g., games).


Research on children's use of wearables for fitness remains sparse, and existing work involves children using devices designed for adults. For example, Miller and Mynatt developed StepStream \cite{miller2014stepstream}, a social fitness system for middle school students that employs FitBit Zip pedometers. They found that students reported a greater sense of enjoyment around fitness, and that the least-active students increased their daily activity. Similarly, ThinkActive \cite{garbett2018thinkactive} has primary students wear a pedometer-based activity tracker to encourage reflection on personal activity. 
\highlight{Both systems, however, provide children with limited access to their data. This work investigates the potentials and challenges of the wearables designed explicitly for children.}


Researchers have also investigated the potential for wearables to support children with neurodevelopmental disorders \cite{american2015neurodevelopmental} such as attention-deficit/ hyperactivity disorder (ADHD) and autism spectrum disorder (ASD). In their review of ASD-support technology for children, Sharmin and colleagues \cite{sharmin2018research} found that wearables were used to enhance social interaction and teach important life skills. More recently, Cibrian and colleagues \cite{cibrian2020supporting} explore the challenges in designing wearables supporting the self-regulation skills of children. They highlight the need for engaging both children and caregivers in setting goals and rewarding behaviors as a means of balancing co-regulation and the long-term goal of self-regulation.  
\highlight{This work investigates the functionality of commercially available wearables for kids to shed light on how these devices can be leveraged to address this need.}

\subsection{Child Surveillance Technology}
Other research on children's wearables concentrates on Child Surveillance Technologies (CSTs) and Child GPS Technologies (CGTs). Marx and Steeves \cite{marx2010beginning} argue that the surveillance of children begins early in life with infant monitoring and nanny cams, and continues in the form of GPS tracking devices once children are old enough to leave the home. 
Infant monitoring technologies (e.g., \cite{owlet, monbaby}) are increasingly common, contributing to what Leaver \cite{leaver2017intimate} calls the normalization of ``intimate surveillance,'' which is defined as the ``purposeful and routinely well-intentioned surveillance of young people'' who have little or no agency to resist. While infant monitors are advertised as a way to provide parents with peace of mind, researchers discovered that the use of a baby wearable technology in the home aggravated parental anxiety and interfered with the social act of parenting \cite{wang2017quantified}. 

Many other researchers have examined perceptions and usage of CGTs in families. Outside the home, CGTs are the predominant method of surveillance. Bettany and Kerrane \cite{bettany2016socio} examine the controversy surrounding CGTs and their usage as mediators in parent-child relationships, as well as the implications for children's welfare and agency. 
Vasalou and colleagues \cite{vasalou2012case} found that parents who favor location tracking feel that tracking technologies provide security, alleviate anxiety, and reduce uncertainty, while parents who oppose location tracking place more value on familial trust and children's self-direction. They discovered that only a minority of parents \highlight{are} in favor of location tracking. 
Boesen and colleagues \cite{boesen2010domestic} found that while location tracking can assist in digital nurturing \cite{rode2010roles}, it also has the potential to undermine trust between parents and children by removing opportunities for trust-building encounters. Ferron and colleagues \cite{ferron2019walk} propose proximity detection devices as a compromise between surveillance and trust, since they support parents' goals of protecting children while supporting their autonomy. J{\o}rgensen and colleagues \cite{jorgensen2016monitoring} shift focus from location tracking to activity monitoring technologies, and found that their usage reduces voluntary information disclosure from children and negatively influences trust in the parent-child relationship. While previous work addresses parental attitudes towards CSTs \& CGTs and the impact of their usage on trust, it is of growing interest to investigate the potential for children's wearables beyond their capability to serve as surveillance tools. 
More recently, Kuzminykh and Lank situate the usage of CGTs within the context of parents' information needs and investigate the types of information that parents seek about their children, their motivations, and potential uses for this information \cite{kuzminykh2019much}. We draw upon this work to understand the purpose and potential usage of features in children's wearables.

\subsection{Family-oriented Information Systems}
One promising research avenue is using technology to promote healthy behaviors in a family context: called ``family informatics,'' \cite{pina2017personal} these technologies enable family members to share and monitor family-oriented activities, as well as health and parenting goals. \highlight{In addition to supporting communication and connectedness (e.g.,~\cite{inkpen2013experiences2go, heshmat2017connecting, tibau2019familysong, forghani2018g2g, yarosh2011mediated}), researchers have explored the potential for family-oriented information systems to promote health and wellness in a family context.} Grimes and colleagues~\cite{grimes2009toward} found that the \textit{collaborative completion} of and \textit{collaborative reflection} on health data can support deeper reflection about health behaviors within the family. Our analysis of design opportunities for kids' wearables is motivated by research on the benefits of family-level tracking for improving children's health. Some family trackers are used by the whole family, as health activities are more sustainable when the entire family is involved~\cite{pina2017personal}. Mobile applications such as Snack Buddy~\cite{schaefbauer2015snack} and TableChat~\cite{lukoff2018tablechat} aim to encourage healthy food choices within families by creating transparency between family members' eating habits. 


Parents can take a large part in monitoring their children's health, guiding them until they gain the ability to self-track. Tools such as Baby Steps~\cite{kientz2009baby} and Estrellita~\cite{hayes2014estrellita} support parents in tracking health data for their infants and young children. 
Devices like WAKEY~\cite{chan2017wakey} and MAMAS~\cite{jo2020mamas} center on helping parents shape their children's habits, allowing the children to take partial responsibility in tracking in order to self-regulate their own routines. For the promotion of physical activity, Saksono and colleagues~\cite{saksono2015spaceship} developed Spaceship Launch, a collaborative exergame that employs activity trackers worn by both parents and children. In later work, Saksono and colleagues~\cite{saksono2018family} noted that research on tools for physical activity promotion in a family context has been limited. \highlight{Recently, Oyg{\"u}r and colleagues~\cite{oygur2020raising} investigated parents' and children's collaborative use of kids' activity trackers. They found that parents used activity trackers to motivate their children to be more active, monitor their children's health and wellness, and teach their children responsibility and independence. These studies motivate our work to evaluate how existing kids' wearables meet parents' information needs for monitoring their children's health, adding to a growing body of literature in HCI that investigates the potential for technology-based, family-oriented approaches to improving children’s health and well-being.}

\section{methods}
\label{sec:methods}
\highlight{Our goal was to synthesize the technological capabilities of commercial kids' wearables and assess the extent to which existing capabilities meet parents' information needs. To that end, we collected commercial devices found online and evaluated their functionality using a combination of affinity analysis and deductive analysis. }



\subsection{Data Collection}
\begin{figure}[h]
    \centering
    \includegraphics[width=.4\linewidth]{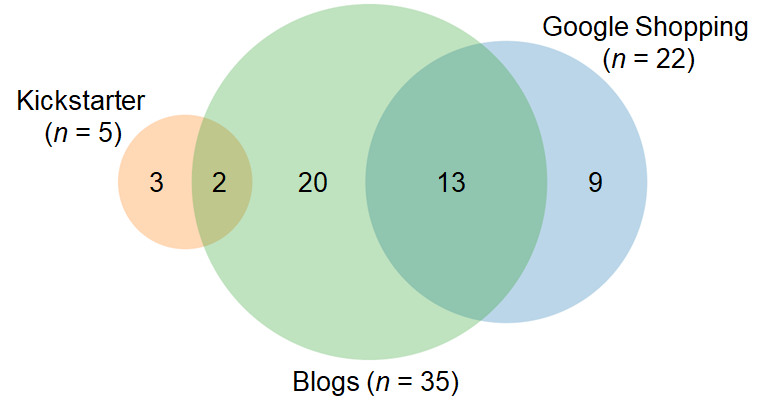}
    \vspace{-2mm}
    \caption{Sources of the 47 devices we found online for analysis.}
    \label{fig:devices}
\end{figure}

We collected a total of 47 devices found online (22 from Google Shopping, 5 from Kickstarter, 35 from parenting blog posts) from February to March 2020 (see ~\autoref{fig:devices}). We ended our data collection once we reached data saturation and encountered mostly duplicates in our search process.
We included devices that (1) were marketed as a children's device; (2) targeted children of ages 3--12; (3) were available (at the time of the data collection); and (4) had a product site available ensuring that we had access to all device details. The level of detail on product websites varied, and during our analysis, a few of the devices became unavailable or discontinued. Because we relied on product descriptions, we were unable to investigate the usability of these devices (e.g., battery life) and explore possible features that were not advertised online. However, we were able to collect rich data about each device's features. To find these devices, we used queries containing the search term ``kids'' and one or more of the following: ``wearable,'' ``tracker,'' and ``smartwatch.'' Example queries include ``kids'' \& ``wearable'' and ``kids'' \& ``tracker.'' We chose to use ``kids'' rather than ``children'' because the former term appears to be more common in marketing, although they are often used interchangeably.

\highlight{Initially, we considered Amazon as a primary online source. However, we found that many products on Amazon did not provide sufficient detail, such as product sites and information about device functionality. On the other hand, the results from Google Shopping provided richer information than those from Amazon, and thus we decided to use Google Shopping to search for devices.} We evaluated up until the fifth page of the results (where each page presents 40 items). 
Items on Google Shopping are not necessarily unique, resulting in a substantial number of duplicate devices. Therefore, instead of recording all items for a search term and then excluding devices, we viewed and included devices as we searched. For each device, if it had not already been recorded, we followed the link to the device. If the device was being sold from a third party seller, we searched the name of the product in Google to find the product site. By the third or fourth page of Google Shopping, most of the devices either did not fit our inclusion criteria, or were duplicates of previous devices. However, we searched up to the fifth page as a precaution. Given the frequency of duplicates and number of listings without product sites on Google Shopping, we turned to other platforms such as Kickstarter and parenting, lifestyle, and tech blogs, which were identified through Google searches using the queries mentioned above. Through our searches on Google, we found 11 blog posts on wearable devices for children \cite{maslakovic2019, maslakovic2019_2, freydkin2020, butcher2020, archambault2020, findmykids2020, turner2020, allison2019, forbes2019, rear2020, snap2020}. \highlight{Examining parenting blogs ensured the inclusion of most popular devices, and we found that many devices appeared repeatedly across blogs as well as in Google Shopping.} 


\highlight{To elicit product features and technical specifications, we collected product descriptions available on Google Shopping and Kickstarter pages, stand-alone product websites, and user manuals (when available). }
For each device, we recorded all of the features with as much granularity as possible. For example, we recorded unidirectional calls with and without auto-pickup as separate ones. 
Features include various methods of communication (e.g., calls, text messages), device settings (e.g., parental restriction of incoming calls), fitness tracking capabilities (e.g., step count, movement), and other dimensions, resulting in 57 unique features. In addition, we captured each device's product name, cost, form factor, target user groups, and method of data access. 



\subsection{Data Analysis}
We first performed affinity analysis~\cite{hartson2012ux} on the data. After collecting all the features for each device, we grouped similar features into representative categories. \highlight{All four researchers met and engaged in this process iteratively and refined initial categories (e.g., ``parent-set contacts'' and ``call firewalls'') into higher-level sub-themes (e.g., ``parental restrictions on incoming and outgoing calls'').}

Next, we performed deductive analysis to contextualize the purpose of the functionality we recorded and further categorize the data. Since many of the devices were advertised to parents and contained features for their benefit (e.g., setting virtual GPS boundaries), we aimed to understand parents' needs and information usage. Kuzminykh and Lank present a categorization of parents' information needs for managing their children, and find that parents are primarily interested in routine information, health information, and social-emotional information \cite{kuzminykh2019much}. Routine information includes sub-themes of sleep, food, physical activity, and location and safety, while health information encompasses general well-being and special situations. Social-emotional information pertains to mood, personal connection, and getting to know and understand a child. While these categories provide a solid foundation, they do not capture all of the features present in the devices. To that end, we adopted some of the themes and sub-themes of Kuzminykh and Lank's categorization and re-categorized them based on the data. For example, ``location and safety'' is a sub-theme under routine information, but emergency and location tracking features were very prominent in the data. Furthermore, location is often a component of safety measures in the wearables we analyzed (e.g., the child presses an SOS button, which calls the parent and sends the child's location). Therefore we elevated ``safety'' to be a higher-level theme and included ``location'' as a sub-theme. We also re-categorized sleep and daily activity as health information rather than routine information, given the impact of sleep and activity on children's health. 

Finally, we created new categories to fully capture the data. For example, Kuzminykh and Lank focus on what information parents want to obtain from their children, but not how they want their children to act or behave. Rather than focusing solely on parents' information needs, we call attention to how wearable devices can be used in caring for children. We created a new category, ``habit formation,'' which encompasses features that may help children develop habits like completing their chores.

\section{Findings}
\label{sec:findings}
Our goal was to investigate the functionality of children’s wearables and classify how existing features support parents’ information needs. We found that wearables for children are complex devices that assist parents in communicating with their children and ensuring their safety, as well as in managing their children’s health and habits. We also identified features that focus on children’s interests rather than parents’ needs. An overview of the devices we analyzed is shown in \autoref{tab:devices}.

\begin{table}[t]
    \centering
    \footnotesize
    \begin{tabular}{@{}lrccccllrccccl@{}}
    \cmidrule(r){1-7} \cmidrule(l){8-14}
        \textbf{Device} & \textbf{Price} & \textbf{S} & \textbf{C} & \textbf{H} & \textbf{N} & &
        \textbf{Device} & \textbf{Price} & \textbf{S} & \textbf{C} & \textbf{H} & \textbf{N} &
        \\
    \cmidrule(r){1-7} \cmidrule(l){8-14}
        Tick Talk\citew{ticktalk}	& \$179.99	& \cellcolor{safety}4	& \cellcolor{connect}2	&	& \cellcolor{nurturing}1 &
        & VTech Kidizoom DX2\citew{vtechkidizoom}	& \$46.49	&	&	& \cellcolor{health}\highlight{2} & & 
        \\
        V-Kids Watch by Vodafone\citew{vkidsbyvodafone} 	& \$159.54	& \cellcolor{safety}5	& \cellcolor{connect}3	&	& \cellcolor{nurturing}2 &
        & UNICEF Kid Power Band\citew{unicefkidpowerband}	& \$39.99	&	&	& \cellcolor{health}3 & & 
        \\
        POMO Waffle\citew{pomowaffle}	& \$189.00	& \cellcolor{safety}6	& \cellcolor{connect}3	& \cellcolor{health}3	& \cellcolor{nurturing}2 & 
        & Smart Tracker by Yepzon\citew{smarttrackerbyyepzon}	& \$151.54	& \cellcolor{safety}4	&	& \cellcolor{health}1 & & 
        \\
        Verizon Gizmowatch\citew{verizongizmowatch2}	& \$179.99	& \cellcolor{safety}3	& \cellcolor{connect}2	& \cellcolor{health}1	& \cellcolor{nurturing}1 &
        & Kiddo\citew{kiddo}	& \$79.00	& \cellcolor{safety}1	&	& \cellcolor{health}2 & & 
        \\
        DokiPal\citew{dokipal}	& \$179.00	& \cellcolor{safety}5	& \cellcolor{connect}4	& \cellcolor{health}3	& \cellcolor{nurturing}1 &
        & Watchitude Move2\citew{watchitudemove2}	& \$60.00	& \cellcolor{safety}1	& &	\cellcolor{health}3 & &
        \\
        myFirst Fone\citew{myfirstfone}	& \$149.00	& \cellcolor{safety}6	& \cellcolor{connect}3	& \cellcolor{health}3	& \cellcolor{nurturing}1 &
        & TIMEX FamilyConnect\citew{timex}	& \$192.00	& \cellcolor{safety}4	& 3 \cellcolor{connect}& & &
        \\
        Ojoy A1\citew{ojoya1}	& \$149.00	& \cellcolor{safety}1 & \cellcolor{connect}4	& \cellcolor{health}5	& \cellcolor{nurturing}1 &
        & Dyno Smartwatch\citew{dyno}	& \$149.99	& \cellcolor{safety}3	& \cellcolor{connect}3 & & &
        \\
        Wizard Watch\citew{wizardwatch}	& \$119.95	& \cellcolor{safety}5	& \cellcolor{connect}2	& \cellcolor{health}1	& \cellcolor{nurturing}2 &
        & Amber Alert GPS\citew{amberalertgps}	& \$135.00	& \cellcolor{safety}6	& \cellcolor{connect}1 & & & 
        \\
        Octopus by Joy\citew{octopusbyjoy}	& \$59.99	& \cellcolor{safety}1	& \cellcolor{connect}1	& \cellcolor{health}1	& \cellcolor{nurturing}1 &
        & KidsConnect GPS Tracker Phone\citew{kidsconnectgpstracker}	& \$129.95	& \cellcolor{safety}5	& \cellcolor{connect}2 & & & 
        \\
        Smart watch For Kids\citew{smartwatchlbs}	& \$57.97	& \cellcolor{safety}4	& \cellcolor{connect}3	& \cellcolor{health}1	& \cellcolor{nurturing}1 &
        & My Gator Watch\citew{mygatorwatch}	& \$117.00	& \cellcolor{safety}1	& \cellcolor{connect}2 & & &
        \\
        OwlCole GPS SmartWatch\citew{owlcole}	& \$27.99	& \cellcolor{safety}4	& \cellcolor{connect}3	& \cellcolor{health}1	& \cellcolor{nurturing}1 &
        & FindMyKids Watch\citew{}	& \$79.99	& \cellcolor{safety}5	& \cellcolor{connect}2 & & &
        \\
        Novus\citew{novus}	& \$199.00	& \cellcolor{safety}4	& \cellcolor{connect}3	& \cellcolor{health}2 & &
        & Weenect Kids\citew{weenectkids} &	\$59.08	& \cellcolor{safety}\highlight{6}	& \cellcolor{connect}1 & & &
        \\
        Xplora 2\citew{}	& \$169.00	& \cellcolor{safety}2	& \cellcolor{connect}2	& \cellcolor{health}1 & &
        & Relay\citew{relay}	& \$49.99	& \cellcolor{safety}3	& \cellcolor{connect}1 & & &
        \\
        Omate X Nanoblock\citew{omatexnanoblock}	& \$149.00	& \cellcolor{safety}2	& \cellcolor{connect}2	& \cellcolor{health}1 & &
        & PocketFinder+ Personal Tracker\citew{pocketfinder}	& \$159.00	& \cellcolor{safety}4 & & & &
        \\
        AngelSense GPS Tracker\citew{angelsense}	& \$99.00	& \cellcolor{safety}5	& \cellcolor{connect}2	& \cellcolor{health}1 & &
        & Trax\citew{trax}	& \$139.00	& \cellcolor{safety}6 & & & &
        \\
        Lil Tracker\citew{liltracker} & \$69.00	& \cellcolor{safety}5	& \cellcolor{connect}3	& \cellcolor{health}1 & &
        & Jiobit\citew{jiobit}	& \$129.99	& \cellcolor{safety}3 & & & &
        \\
        Anti Lost Child Wristwatch\citew{antilostchildwristwatch}	& \$29.99	& \cellcolor{safety}4 & \cellcolor{connect}3	& \cellcolor{health}1 & &
        & Samsung Smartthings Tracker\citew{samsungsmartthings}	& \$99.99	& \cellcolor{safety}3 & & & &
        \\
        Revibe Connect\citew{revibeconnect}	& \$119.95	&	& \cellcolor{connect}1	& \cellcolor{health}1	& \cellcolor{nurturing}1 &
        & Gego Universal Tracker\citew{gego}	& \$99.95	& \cellcolor{safety}3 & & & &
        \\
        Garmin Vivofit Jr 2\citew{garminvivofitjr2}	& \$69.99	&	& \cellcolor{connect}1	& \cellcolor{health}7	& \cellcolor{nurturing}2 &
        & Tracki 2020\citew{tracki} & \$49.00	& \cellcolor{safety}3 & & & &
        \\
        Kurio\citew{kurio}	& \$65.00	&	& \cellcolor{connect}2	& \cellcolor{health}2	& \cellcolor{nurturing}1 &
        & Girafus Guardian Angel Child\citew{girafusguardian} & \$39.99	& \cellcolor{safety}3 & & & &
        \\
        Fitbit Ace 2\citew{fitbitace2}	& \$49.95	&	&	& \cellcolor{health}4	& \cellcolor{nurturing}2 &
        & Buddy Tag\citew{buddytag}	& \$39.99	& \cellcolor{safety}3 & & & &
        \\
        Sqord Activity Pod\citew{sqord}	& \$40.99	&	&	& \cellcolor{health}4	& \cellcolor{nurturing}1 &
        & Mommy I'm Here Child Locator\citew{mommyimhere}	& \$29.95	& \cellcolor{safety}1 & & & &
        \\
        iTouch PlayZoom Smartwatch\citew{itouchplayzoom}	& \$29.99	&	&	& \cellcolor{health}1	& \cellcolor{nurturing}1 &
        & Smart B'ZT\citew{smartbzt} & Price varies & \cellcolor{safety}1 & & & &
        \\
        LeapFrog LeapBand\citew{leapfrogleapband}	& \$29.99	&	&	& \cellcolor{health}3	& \cellcolor{nurturing}2 &
        &
        \\
    \cmidrule(r){1-7}     \cmidrule(l){8-14}
    \end{tabular}
    \vspace{1mm}
    \caption{The name and price of 47 wearable devices we analyzed, along with the number of supported features grouped by their main categories: S (Safety); C (Connectedness); H (Health); and N (Nurturing Good Behaviors and Habits).}
    \vspace{-3mm}
    \label{tab:devices}
\end{table}

\subsection{Children's Safety}

Many features embedded in wearable devices aim to offer parents \textit{peace of mind} about their children's safety, mainly by providing information about children's whereabouts, enabling rapid communication in case of emergency, and offering on-demand video and audio surveillance. This scheme is exemplified in Wizard Watch's\citew{wizardwatch} advertising phrase: \textit{``Gain peace of mind by knowing exactly where your child is, how to reach them... and knowing they can alert you if they are in trouble, every moment of the day!''} In this section, we detail the specific features that support children's safety (\autoref{tab:safety}).

\begin{table}[t]
\begin{tabular}{p{4cm}p{9.2cm}r}
\hline
\textbf{Features} & \textbf{Description (number of devices)} & \textbf{\makecell{\# Devices \\ (\textit{N} = 47)}} \\ \hline
Access to location (GPS)                & Parents can view children's real-time location (32) and location history (12)                                & 32                         \\ \hline
Set geofence                            & Parents can set virtual geographic boundaries and receive alerts when children enter or exit specified zones (e.g., school)                                                                                                                   & 27                         \\ \hline
SOS call                                & Children can call/alert parents in case of emergency, normally by pressing a designated button                                                                                                                & 27                        \\ \hline
Device-level controls and alerts                     & Parents have remote control over device settings (9) (e.g., set quiet mode) and are alerted to device states such as inactivity (2) and removal (2)                     & 11                         \\ \hline
Accessible by multiple caregivers       & Other caregivers (e.g., grandparents) can access the device's data                                                                                                                      & 10                         \\ \hline
Audio \& video surveillance                       & Parents can access the device's microphone for voice monitoring (10) or the camera for video monitoring (2)                                                        & 10                         \\ \hline
Proximity detection & Parents receive an alert when children move farther than a set distance away from them                                                       & 6                         \\ \hline
Send emergency info                     & In case of an emergency, the device automatically sends the child's GPS location (4) or voice recording of the child's surroundings (2) to emergency contacts & 5                         \\ \hline
In transit alert                        & Parents are alerted to transit incidents, such as high speeds (2) or collisions (1)                                             & 3                          \\ \hline
Send directions to child                & Parents send directions to let children self-navigate                                                    & 2                          \\ \hline
\highlight{Augmented reality}                       & \highlight{Device annotates children's location and distance on a real-world view seen through the parent's camera} & \highlight{2} \\ \hline
Store medical info                      & Device stores the child's medical information                                   & 1                          \\ \hline
\end{tabular}
\vspace{1mm}
\caption{Safety features concentrate on parents' ability to monitor their children and communicate with them in times of emergency. Monitoring emphasizes the collection of information related to children's whereabouts, and a substantial portion of devices also provide on-demand video and audio surveillance.}
\vspace{-3mm}
\label{tab:safety}
\end{table}


\subsubsection{Whereabouts}

Our results show that a majority of devices support safety features, with an emphasis on information about children's whereabouts. This information is typically viewed on a partner mobile application belonging to the parent. Parental access to children's real-time GPS location is the most common feature overall, appearing in 68\% of devices (\textit{n} = 32), and nearly all devices with GPS tracking \highlight{(27/32)} also enable geofencing (i.e., setting virtual geographic boundaries). For some parents, continuous real-time information about children's movements can be seen as a way to reduce uncertainty, alleviate personal anxiety, and provide a sense of security~\cite{vasalou2012case}. In some cases, location tracking can afford more than momentary relief: 26\% of devices (\textit{n} = 12) provide location history, which gives parents more insight into how and where children spent their time. 



Access to location information can be useful in both routine and emergency situations. With geofences, parents can verify that routine activities were performed without an issue. For example, a parent can set geofences around their home and the child’s school and receive daily alerts to confirm that the child completed their journey. In case of an emergency, accessing real-time location allows parents to come to their children's aid. To continue with the previous example, if the child gets lost on their way home from school, triggering an alert, then the parent can use location information to plan their response (e.g., calling to give the child directions or going to pick them up). This functionality is in line with Kuzminykh \& Lank’s finding that one of the primary use-cases for parental information is performing emergency activities \cite{kuzminykh2019much}.

Although GPS is the dominant form of location tracking, six devices use proximity detection to inform parents about their children's whereabouts. For example, Buddy Tag\citew{buddytag} will send an out of range alert to a parent's phone when their child moves farther than 80 to 120 feet away in an outdoor space. Another device, Mommy I'm Here Child Locator\citew{mommyimhere}, sounds an alarm on the child's device to help parents locate them more easily in public. \highlight{Two devices, Weenect Kids\citew{weenectkids} and Trax\citew{trax}, take a different approach to proximity detection by incorporating augmented reality. This functionality makes it easier for parents to identify where their children are in crowds and how far away they are by overlaying children's location and distance on a real-world view using the phone's camera.} A key difference between GPS tracking and proximity detection is that the former is used when children and parents are separated, while proximity detection can only be used when children and parents are together.

\subsubsection{Alerts in Emergency Situations}
In addition to features locating kids' whereabouts, most devices have at least one feature designed to notify parents of emergency situations, or events that occur outside of children's normal routines. While Kuzminykh \& Lank classified special situations as a subtheme of health information, we found that health emergencies (e.g., fever) were not a primary focus. Instead, the information collected in these situations emphasizes a child's immediate safety, which may be linked to whereabouts. One widespread feature that supports emergency activities is SOS buttons, appearing in 57\% of devices (\textit{n} = 27). By pressing a designated button, children can discreetly and easily call their parents in an emergency (e.g., getting lost on their way to school). In four devices, pressing the SOS button will also send the child’s location to the parent, suggesting that location and communication are closely linked with ideas of safety. Another dimension of SOS functionality is that some devices alert multiple emergency contacts, rather than just the parent. For example, KidsConnect GPS Tracker Phone\citew{kidsconnectgpstracker} will text the child's GPS location to three emergency contacts, and then call until one of them answers. The responsibility to ensure a child’s safety and respond to emergency situations typically does not rest solely on one parent, and some devices aim to distribute this responsibility among trusted adults in a child's life. 
\highlight{Ten devices} allow multiple caregivers to access the data collected about a child, as seen in \autoref{tab:safety}, with a focus on the sharing of location information. Jiobit\citew{jiobit} refers to these other caregivers as part of the ``village’’ it takes to raise a child, including nannies, co-parents, grandparents, or any other trusted individuals. 

Other features for emergency situations include device-level alerts such as device inactivity alerts and device removal alerts, as well as in transit alerts for speed and collisions, as seen in \autoref{tab:safety}. These features can inform parents when their children are in immediate danger or might require assistance. Device inactivity and removal alerts are somewhat distinct in that they can notify parents of potential future threats rather than immediate danger (i.e., the parent will not have the information they need to respond to an emergency if the child is not wearing the device). AngelSense\citew{angelsense}, a wearable designed for people with special needs, goes a step further by allowing device removal only with a special parent key.

\subsubsection{On-Demand Video and Audio Surveillance}
Children's wearables are approaching safety in unprecedented ways beyond the collection of location information. A 
large share of devices allow parents to remotely monitor their children without 
\highlight{notification} as a means of ensuring safety or security. Twenty one percent of devices (\textit{n} = 10) allow parents to remotely listen to the children's surroundings using a built-in microphone, and two devices even enable remote camera monitoring to view children's surroundings. Find My Kids GPS-watch\citew{findmykidsgps} describes audio monitoring as a way to \textit{``always be aware of the sounds surrounding your child. Know immediately if they're being bullied or going through a rough time, or potentially in danger.''} Similar to real-time GPS location, remote monitoring is marketed as a way to check on children throughout the day and have \textit{peace of mind} about their well-being. Information captured through audio can also be used in emergency situations. For example, two devices send parents an audio recording of the child's surroundings when the child presses the SOS button. These recordings can provide parents with information about the type of emergency, other involved parties, and their child's status, which kids might not be able to communicate well.

\subsection{Connectedness}
Another primary purpose of wearables is to facilitate communication between parents and children. Some devices also aim to connect children with their peers. In this section, we highlight the complexities of children's wearables as communication devices (see \autoref{tab:communication} for an overview).

\subsubsection{Connecting with Parents}
Our data suggest that communication features in children's devices are sophisticated and varied with respect to the modality of communication, the level of control to initiate and decline communication, and parties allowed to communicate through the device. Modalities of communication include voice, video, text, and vibrations. Voice calls are the most common communication medium (\textit{n} = 21), followed by text messages (\textit{n} = 14). The power to initiate and decline communication adds another layer of complexity: communication can be unidirectional (e.g., from parent to child) or bidirectional, and one device enables auto-pickup (i.e., a child cannot refuse a call from their parent). By enabling fine-grained control, children's wearables position themselves as safe alternatives to phones that can meet parents' specific needs. For example, My Gator Watch~\citew{mygatorwatch} advertises itself as \textit{``For 5-12 years old kids who are too young for mobile phones.''} Further, Weenect GPS tracker~\citew{weenectkids} promises parents, \textit{``they} [children] \textit{can call you but not send messages to their friends. If they don't answer, no worries, you know the GPS location of the tracker. The advantages of the mobile phone without the disadvantages.''} Children's wearables afford parents with the ability to communicate with their children when necessary, while also protecting them from the perceived dangers of mobile phones. Eleven devices enable parental restrictions on incoming and outgoing calls, which include pre-programming approved phone numbers for outgoing calls and blocking incoming calls. By providing parents with contingency plans (e.g., auto-pickup and access to location), wearables also provide parents with security in case children do not respond to communication requests.

\subsubsection{Connecting with Friends}
While communication features center on connecting children with their parents, some devices draw attention to children's ability to connect with their peers. In total, eight devices let children add friends with the same watch, though the supported methods of communication vary. Some devices allow children to call their friends (e.g., POMO Waffle\citew{pomowaffle} and DokiPal\citew{dokipal}), while others enable voice messages and emojis between friends (e.g., V-Kids Watch by Vodafone\citew{vkidsbyvodafone} and Ojoy A1\citew{ojoya1}). Devices market this functionality not only as a way to motivate children to use the device (e.g., ``more fun with friends''\citew{dokipal}), but also as an opportunity to help children learn how to communicate better (e.g., ``[improve] their social skills''\citew{pomowaffle}, ``broaden your child’s interpersonal skills''\citew{myfirstfone}). Additionally, connecting with friends can take place in the context of activity competitions (e.g., highest step count), in which children can engage with their friends and classmates. However, there does not appear to be a direct relationship between the ability to communicate with peers and the presence of activity competitions (i.e., most devices have either one feature or the other). Regardless, both features support children in interacting with their peers and exercising their social skills.



\begin{table}[]
\begin{tabular}{p{4.4cm}p{8.9cm}r}
\hline
\textbf{Feature}                                     & \textbf{Description (number of devices)}                                                                       & \multicolumn{1}{l}{\textbf{\makecell{\# Devices \\ (N = 47)}}} \\ \hline
Calling                                                & Parents and children can call using voice (21) and video (5)                                                                 & 22                                             \\ \hline
Messaging                                             & Parents and children can message using text (14), voice (10), group messaging (2), and video (1)                              & 19                                             \\ \hline
Parental restrictions on incoming and outgoing calls & Parents can limit communication for children (e.g., setting call firewalls) & 11                                             \\ \hline
Add/ find friends with the same watch                 & Children can add friends with the same device to a contact list and communicate with them                            & 8                                             \\ \hline
One-way communication (parent to child)         
& Only parents can initiate contact with children through haptic feedback (2), calling with auto pickup (1), and texting (1)                       & 3                                             \\ \hline
\end{tabular}
\vspace{1mm}
\caption{Wearables provide various mediums of communication (e.g., voice and video calls, text messages) to connect children with their parents and with their peers. Additional features such as parental restrictions equip parents with more oversight regarding their kids' communication.}
\vspace{-3mm}
\label{tab:communication}
\end{table}


\subsection{Children's Health}

Around half of devices provide parents with health-related information about their children, primarily related to physical activity. A minority of devices provide some other type of health-related information such as sleep, temperature monitoring, and perspiration monitoring. \autoref{tab:health} shows an overview of health features.

\begin{table}[]
\begin{tabular}{p{4.4cm}p{9cm}r}
\hline
\textbf{Features}           & \textbf{Description (number of devices)} & \textbf{\makecell{\# Devices \\ (\textit{N} = 47)}}
\\ \hline
Activity tracking           & Tracks children's daily activity, such as step count (21), calories burned (5), movement (5), and distance (4) & 26                                             \\ \hline
Activity-based competitions                & Compete (e.g., for highest step count) with friends (6) and family (3) & 6                                             \\ \hline
Built-in rewards            & Rewards given by device for completing activity goals (e.g., in-game upgrades) & 5                                             \\ \hline
Parent-set activity goals   & Activity goals parents set for child                                                                                   & 5                                             \\ \hline
Activity games              & Games to promote health and activity                                                                                   & \highlight{5}                                              \\ \hline
Built-in activity goals     & Automatic activity goals set by device                                                                                 & 4                                              \\ \hline
Parent-set rewards          & Rewards given by parent for completing activity goals & 4                                              \\ \hline
Sleep tracking              & Tracking daily sleep of child                                                                                          & 4                                              \\ \hline
Cardiac \& body measurement & Monitoring body states such as temperature and heart rate               & 2                                              \\ \hline
\end{tabular}
\vspace{1mm}
\caption{Health features focus on activity tracking, as well as promoting physical activity through competitions, goals, and rewards. A few devices track other types of health-related information such as sleep and cardiac and body measurement.}
\vspace{-5mm}
\label{tab:health}
\end{table}

\subsubsection{Activity tracking}
We observe that 55\% of devices provide support for activity tracking, with a focus on step count (\textit{n} = 21). To a lesser extent, some devices track calories burned (\textit{n} = 5), movement (\textit{n} = 5), and distance (\textit{n} = 4). Unlike activity trackers for adults, which emphasize caloric expenditure and heart rate in pursuit of specific activity goals (e.g., losing weight), wearables for children are more focused on general physical activity promotion and habit formation. Step count is typically displayed on the device as a standalone number, and some devices also situate children's activity in the context of their activity goals (e.g., Garmin Vivofit Jr 2\citew{garminvivofitjr2} shows a radial progress bar for active minutes). For more detailed information (e.g., aggregate data over a week-long period), children need to use their parents' phones. Other devices shift the focus away from counting steps by instead converting movement into points (e.g., Octopus by Joy\citew{octopusbyjoy}). Sqord\citew{sqord} argues that by converting movement into points, which can be used for in-game rewards, the device can ``track play'' rather than just steps. 

There are different methods to encourage children to be active: goals, rewards, games, and competitions. Goals and rewards can be built-in to the device or set by parents. As an example of a built-in goal and reward, Ojoy A1\citew{ojoya1} has a target of 4000 steps per day and rewards children with an upgrade to their in-game character if they achieve that target. POMO Waffle\citew{pomowaffle} allows parents to customize activity goals and create real-life rewards with their children. Another way to encourage activity is through games and competition. Four devices offer activity-based games that require children to move (e.g., ``jump like a frog'' in LeapFrog LeapBand\citew{leapfrogleapband}). Access to more games, which are not necessarily activity-based, can also be used as a reward. For example, Garmin Vivofit Jr 2\citew{garminvivofitjr2} makes certain games available to children in the mobile app once they complete 60 minutes of daily activity and incentivizes activity beyond 60 minutes with more opportunities to play. In addition to games and rewards, socially-based activity events encourage children to be active. Six devices let children engage in activity competitions with friends, and three devices allow for family-based activity competitions. By framing exercise as an enjoyable social practice, children’s wearables can help build social support for physical activity and motivate children to increase their activity levels while having fun. Family practices around exercise also have a significant impact on children’s activity levels, and competitions can engage the entire family in building healthier habits. 

\subsubsection{Other Metrics of Health}
Although activity tracking is the main focus of children’s wearables, our data show that other types of health-related information may be of interest to parents. Some devices track sleep (\textit{n} = 4), temperature (\textit{n} = 2), and heart rate (\textit{n} = 1). Details on sleep tracking are sparse, with little additional information provided beyond time spent asleep. One device, Fitbit Ace 2\citew{fitbitace2}, analyzes sleep quality using the number of times awakened and restless. While sleep is an important topic in terms of habit formation (adhering to a bedtime schedule), it does not appear to be a major health concern. Only one device, Kiddo\citew{kiddo}, tracks sleep and key vital signs (temperature, heart rate, skin temperature, and perspiration) to assist parents in communicating with healthcare providers. Children without specific health conditions usually do not have specific health concerns \cite{pina2017personal}, which may explain the shortage of devices with cardiac and body measurement capabilities.

\subsection{Nurturing Good Behaviors and Habits}

In this section, we draw attention to features that assist parents in helping their children form desirable habits or behaviors (see \autoref{tab:habits} for an overview). These habits are not restricted to a particular domain; rather, they encourage children to autonomously engage in a variety of behaviors, such as completing their chores and staying focused on relevant tasks. These features support varying degrees of parental involvement. 

\begin{table}[]
\begin{tabular}{p{4cm}p{9.2cm}r}
\hline
\textbf{Features} & \textbf{Description (number of devices)} & \textbf{\makecell{\# Devices \\ (\textit{N} = 47)}} 
\\ \hline
Parent-set scheduled events and alerts & Parents can set or send children tasks and reminders                                                                                                          & 12                                             \\ \hline

Educational games                       & Games that promote learning                                                                                                                                   & 4                                              \\ \hline

Children-specific view in mobile app    & Mobile app has a view designed for children to access their own data      & 3                                              \\ \hline
Send directions to child                & Parents can let child self-navigate by sending directions to device & 2                                              \\ \hline
Focus rate tracking                     & Parents can track their child's focus rate, attention span and other data points to measure their on-task behavior & 1                                          \\ \hline
\end{tabular}
\vspace{1mm}
\caption{Some devices promote good habits and behaviors such as adhering to a schedule and developing a regular routine, mainly by allowing parents to schedule events and reminders for their children.}
\vspace{-3mm}
\label{tab:habits}
\end{table}

The most common feature for habit formation is parent-set scheduled events and alerts, which are present in 26\% of devices (\textit{n} = 12). Parents can schedule tasks for their children, such as brushing their teeth or completing their homework, to send reminders. 
The objective is to train children to adhere to schedules and develop a regular routine, so that at some point they do not need to be reminded. Octopus by Joy\citew{octopusbyjoy}, for example, allows parents to create a visual schedule for their children through the usage of icons, which it says \textit{``empowers kids by teaching good habits and the concept of time.''} Revibe Connect\citew{revibeconnect}, which is designed for children with ADHD, uses vibration reminders to signal the start and end of work intervals (15 minutes of work, 5 minutes of break) to teach children to work independently and \textit{``become aware of their behavior and get back on-task.''} Habit formation can also be supported by rewards, which are either set by parents or built-in to the device. With Garmin Vivofit Jr 2\citew{garminvivofitjr2}, parents can assign virtual coin values to chores and let children redeem them for agreed-upon rewards in real life. By involving both parents and children in this process, children’s wearables can simultaneously encourage healthy habits and facilitate connectedness between parents and children. Other devices favor a more restrictive approach through remote parental control of device settings (\textit{n} = 9), which include turning on quiet mode, deactivating the device, and disabling games. Through restricting device functionality, parents can encourage children to stay focused on the task at hand and limit their screen time. 


\subsection{Designing for Children's Interests}

Although the information collected from wearables is mainly meant for parents' consumption, the devices are worn and used in practice by children. 
\highlight{Four} features fall outside of the structure for parents' information needs and instead cater to children's interests. These features include physical customization of the tracker (e.g., change the watch face or wristband); in-game customization of the tracker (e.g., choose an avatar); 
a voice assistant; and recreational games (i.e., not educational or activity-based). Twenty one percent of devices (\textit{n} = 10) allow for some type of tracker customization. Previous research showed that customization features in wearable health trackers may increase one's sense of identity, which in turn is associated with favorable attitude, higher exercise intention, and greater sense of attachment towards the tracker \cite{kang2017fostering}. As such, providing customization features could prevent device abandonment and may incentivize children to continue using the wearable and encourage feelings of ownership, as opposed to feeling forced to use the device (e.g., with device removal alerts). 

\section{Discussion and Future Work}
\label{sec:discussion}

The technological landscape of wearable devices for children is diverse. In this section, we discuss the key themes that emerged from our data, situate our findings in the context of parents' information needs, and make design recommendations for children's wearables. We also highlight opportunities for future work, including the integration of collaborative tracking into children's devices.

\subsection{Insights from the Current Landscape of Children's Wearables}

Recent work by Kuzminykh \& Lank \cite{kuzminykh2019much} considers wearables as a means to satisfy parents' information needs (routine information, health information, schooling information, and social-emotional information). We found that there is a mismatch between parents' stated information needs and the information that is collected by children's wearables. Although Kuzminykh \& Lank's \cite{kuzminykh2019much} semi-structured interviews revealed that parents resist continuous monitoring and are disinterested in location information, we found that a majority of devices are built for ongoing parental surveillance as a means of ensuring children's safety. Existing devices focus on the collection of safety information (e.g., real-time GPS location) rather than routine information, and physical activity data instead of general health information. Schooling information was not a focus of the devices we surveyed, and while most devices support parent-child communication, they do not necessarily provide the social-emotional information that parents are interested in. 

Despite wearables' emphasis on monitoring, our findings suggest that these devices have great potential beyond their capacity for surveillance. Close to 40\% of devices are equipped with features to support children in developing good habits and self-regulation skills in the context of daily routine and time-management. Just as adults practice self-tracking to achieve various goals (e.g., improving health and productivity) \cite{choe2014understanding}, children stand to benefit from practicing self-tracking at an early age. Self-tracking can engage children in planning and reflection, which help children practice self-regulatory mechanisms and develop higher-level thinking skills \cite{epstein2003planning}. However, more research is needed to understand how devices can support children in playing the main role in tracking and consuming their own data, rather than acting as passive wearers who collect data for their caregivers. In the following, we discuss the benefits and drawbacks of children’s wearables, and how children can play a more active role in the design and use of these devices.

\subsection{Supporting Kids' Self-tracking}

Unlike wearables for adults, none of the wearables for children that we surveyed are marketed as ``self-tracking'' devices. Instead, the primary goal of children's trackers is parental surveillance. To effectively use wearables in promoting children’s self-regulation, designers and researchers must re-envision children's role in the tracking process and empower them to become active stakeholders. As a starting point, designers should consider features that support children's agency (i.e., sense of control) and motivate them to self-track. \highlight{Self-determination theory has shown that a feeling of agency enhances children's intrinsic motivation to perform target activities (e.g., physical activity \cite{chatzisarantis1997self}) and encourages the development of self-regulation \cite{grolnick1997internalization}. As such, supporting children's sense of autonomy, agency, and empowerment has become a major topic of interest in the research community \cite{kawas2020another}. Recently, Oyg{\"u}r and colleagues \cite{oygur2020raising} highlighted the need for kids' activity trackers that support children's agency. In their analysis of online reviews for nine kids' activity trackers, they found that parents expected trackers to teach their children to become more responsible for their health, daily tasks, and schedule. However, a lack of agency on children's side increased parents' workload and diminished their goal of facilitating children's independence. Recognizing this need, we provide design recommendations for wearables that enhance children's agency, which is not well-reflected in existing commercial wearables for kids.}

While a few devices we analyzed allow children to make cosmetic changes to their devices (e.g., changing the watch face), wearables could provide more meaningful customization options, such as letting children set individualized goals and track behaviors that are important to them. Ananthanarayan and colleagues \cite{ananthanarayan2016craft} explored the benefits of allowing children to craft their own tangible health visualizations based on data collected by wearable trackers, and found that children interpret health and wellness as the ability to perform activities that are important to them. Instead of focusing on traditional metrics of health like adults' wearables (e.g., heart rate), devices should support children's activities of interest and aim to introduce them to the idea of managing health and well-being, as well as help them become self-aware about their own health behaviors. Another important consideration in maintaining children's interest is to make the device fun to use \cite{hutchinson2003technology, ananthanarayan2010health}. Furthermore, rather than simply adopting the feedback mechanisms present in adults' wearables (e.g., graphical representations of activity data), designers should consider feedback that is easier for children to interpret. Toward this new design direction, involving kids in the design process early on can help designers elicit kids' values, motivators, and design insights. HCI researchers developed many design methods and techniques to involve kids during the design process, such as cooperative inquiry~\cite{druin1999cooperative} and co-design using fictional inquiry and comicboarding~\cite{hiniker2017co}, which can be employed to partnering and collaborating with kids.

\subsection{Collaborative Tracking and Reflection}

Given that children cannot be expected to independently engage in self-tracking and reflection, parental involvement is crucial. Collaborative tracking and reflection present an opportunity for parents to teach their children good habits and behaviors, while also encouraging the development of self-regulation. To encourage sustainable, long-term collaborative tracking, scaffolding may be employed accounting for children’s developmental progress and its impact on parent-child tracking dynamics. When children are young, wearables can help parents become involved in the tracking process to assist children in setting goals, interpreting feedback, and reflecting on the data. \highlight{Collective goal-setting can support health reflection within families \cite{colineau2011motivating}.} At this stage, parents might play a more active role in reminding their children to track, and they may even record some data for children if they are unable. As children age and begin to understand the purpose of self-tracking, parents may decrease the amount of control that they have over their children’s tracking behaviors and support children in completing tasks independently. \highlight{For example, children might set their own goals without parental input, but then reflect on their data with parents, who can add new insights and context to children's individually-collected data \cite{grimes2009toward}.} Through this process of collaborative tracking, wearables can create shared experiences for parents and children and facilitate emotional bonding \cite{saksono2015spaceship}. That said, tracking alone does not necessarily lead to deeper reflection as a family, and therefore tracking tools should facilitate discussions around health data to help parents teach their children healthy habits \cite{saksono2019social}.
An exciting avenue for future research is to explore different ways to share tracking responsibilities between parents and children, as well as ways to balance data access and the amount of information shared. While one option is to employ parents as ``guides'' in teaching children how to self-track, another option is to have parents track same or different data alongside their children (e.g., \cite{pina2020dreamcatcher}). 



\subsection{Wearables as an Alternative to Smartphones}

For parents who want to communicate with their children but are worried about mobile and online safety (e.g., \cite{jones2012trends, livingstone2011risks}) and smartphone addiction \cite{haddon2015uk}, wearables present a compelling alternative. Because wearables can serve as children's primary ``smart'' devices, many wearables have communication features that are on par with smartphones (voice \& video calls, and text messaging). At the same time, they provide parents with increased control over their children's communication (e.g., whitelisting contacts) and contingency plans in case of communication failures (e.g., on-demand audio monitoring and location information). However, there are limitations to replacing smartphones with wearables. The small screen and form factor of wearables can make it difficult for children to view their self-tracking data, causing them to rely on parents' devices. For example, several devices provide a ``kid view'' in the mobile app on parents' phones to allow children to access their data. Future work might explore how to turn the need for cross-device collaboration into a pleasurable experience for parents and children (e.g., as an opportunity for reflection and bonding). 

We also highlight opportunities for wearables to serve as more than communication devices by facilitating connectedness between parents and children. While about half of the devices we surveyed allow for some type of communication between parents and children, the presence of communication features alone does not necessarily promote connectedness or address parents' desire for social-emotional information about their children. From a study on communication between working parents and their children in China, Sun and colleagues found that mobile phones, despite being the most common means of communication, were rated by children as the least desirable way for parents to show their love \cite{sun2016seesaw}. In response, they developed e-Seesaw, a tangible awareness system that facilitates connectedness by letting parents and children play together remotely. Given that young children may also struggle to communicate their needs and wants through speech, more research is needed to explore alternative methods of communication between parents and children. Rather than serving as simple communication devices or monitoring tools, children’s wearables have the potential to enrich parent-child communications and thus enhance their relationships.

\subsection{Limitations}
\highlight{We aimed to capture a snapshot of the current landscape of kids’ wearables. Although some research prototypes may be more advanced than commercial kids' wearables, consumer wearables can still provide valuable insight into how these devices are intended to be used in a real-world context, given that research prototypes of kids' wearables are relatively sparse and not widely adopted. This work complements existing research by analyzing the functionality of existing commercial kids' wearables and identifying new directions in design.} 

\highlight{Our findings rely on data collected from product descriptions available on Google Shopping, stand-alone product websites, and user manuals (when available). We acknowledge that these materials may not accurately represent the quality and usability of devices, as well as parents' and children's experiences with the devices. To address this limitation, we referred to Oyg{\"u}r and colleagues' study \cite{oygur2020raising}, which examines parents' perspectives of kids' activity trackers through an analysis of device reviews. Additionally, we relied on Kuzminykh and Lank's work \cite{kuzminykh2019much} on parents' information needs to understand the potential purpose of device functionality. The use of online sources limited our ability to investigate certain features, though we note that this data collection method allowed us to provide broad coverage of a diverse array of devices, which would have been infeasible with in-depth user studies and interviews. 
}

\section{Conclusion}
\label{sec:conclusion}
In this work, we \highlight{surveyed the technological capabilities of commercial kids' wearables and examined how these devices address parents' information needs.} We found that while many wearables emphasize children's safety and ability to communicate with their parents, a significant share are also equipped with features to support their health and the development of good habits and behaviors. Kids' wearables often have rich functionality, offering GPS tracking, activity tracking, and rich communication features. As we rethink the design of kids' wearables, we envision leveraging a combination of self-tracking and collaborative tracking to balance the needs of both parents and children. We believe that self-tracking can help children develop self-regulation skills and an awareness of their own behaviors, while collaborative tracking can facilitate parental involvement and help parents teach their kids good habits. Promising areas for future research include exploring ways to design for supporting kids’ agency and interest, to leverage cross-device collaboration, and to share tracking responsibilities between parents and children. Wearable devices that empower kids and parents as equal stakeholders could promote health and wellness at a family level and help children develop the skills they need to be independent in the long-term, while supporting parents' short-term goals of ensuring their kids' safety and well-being. Our work contributes to the nascent field of family-centered informatics with a focus on the role of wearable devices. 

\section{Selection and Participation of Children}
No children participated in this work.

\bibliographystyle{ACM-Reference-Format}
\bibliography{references}

\bibliographystyleW{ACM-Reference-Format}

\makeatletter
\renewcommand{\@biblabel}[1]{W#1.}
\makeatother
\bibliographyW{wearables}



\end{document}